\documentstyle[12pt]{article}
\nopagebreak

\begin{document}

\centerline{{\Large\bf Quantum interference effects}}

\vskip 0.5cm

\centerline{\Large{\bf in $N^*$ electroproduction and propagation
in nuclei}}

\vskip 1.5cm

\centerline{\large{A. Bianconi, S. Boffi and D.E. Kharzeev}
\footnote{On leave from Moscow University, Moscow, Russian
Federation}}
\vskip 1.0cm

\centerline{\small Dipartimento di Fisica Nucleare e Teorica,
Universit\`a di  Pavia, and}

\centerline{\small Istituto Nazionale di Fisica Nucleare,
Sezione di Pavia, Pavia, Italy}

\vskip 1.5cm

\begin{abstract}
We discuss the unexpected enhancement of the $N^*$
electroproduction on nuclei recently observed  in the
$(\gamma^*,p\pi^-)$ measurement. A  mechanism which is able to
explain this result is proposed. To clarify   the situation, we
suggest to perform a new kind of experiment within well
specified kinematic conditions.
\end{abstract}

\vskip 1.2truecm

Recently a measurement has been reported [1] of the
($\gamma^*,{\pi^-}p$) reaction in the second resonance  region
($W \approx$ 1530 MeV) at different values of the invariant mass
$W$ and of the missing momentum $p_m$.  This provides valuable
information on $N^*$ electroproduction and propagation in
nuclei. In the same work the data have  been compared with
theoretical expectations based on a  DWBA ($\gamma^*,{\pi^-}p$)
calculation. The new result is that, for $W \approx$ 1530 MeV,
at high $p_m$ (222 MeV) the  data to DWBA ratio is 1.9 $\pm$ 0.3
(while at $p_m$ = 56 MeV this ratio is $\approx$ 1). After
discussing the possible causes of this mismatch, the authors of
Ref. [1] suggest that it could be due to Final State
Interactions (FSI) between $N^*$ and  the residual nucleus. They
also stress that at $W \approx M(N^*)$ the data are above the
results of the DWBA calculation, while at $W \approx M(\Delta)$
the trend is opposite. This indicates that FSI of $N^*$ and
$\Delta$  are largely different, owing to their different
quantum numbers.

In this note we propose a mechanism which is in the spirit of the
suggestions of Ref. [1] and can account for an increase  of $N^*$
production in nuclei, critically depending on $p_m$.  We give
numerical estimations of the effect and suggest some possible
experimental  developments.

Due to the nuclear Fermi motion, the kinematics of the
$\gamma^*N$ hard collision is not fully determined, even when
each component of the exchanged four momentum $q^\mu$ is known.
Let $\Phi(k)$ be the amplitude of finding a bound nucleon with a
momentum $k^\mu$;  $s = (q + k)^2 \equiv M^2$ is the squared
center of mass energy  of the $\gamma^*N$ system, and the
squared mass of the baryon state produced after the $\gamma^*$
absorption on the nucleon; $\Phi(k)$ induces an amplitude $f(M)$
for the values of this mass.  So, according to the quantum
superposition principle, many mass eigenstates are {\sl
simultaneously}\/ produced in  a $\gamma^*N$ collision in
nucleus. Mass eigenstates diagonalize the {\sl free}\/
hamiltonian, but do not diagonalize the hamiltonian which
describes the propagation of  a hadronic wave packet inside a
nucleus. In fact the interaction with the nuclear environment
can convert a nucleon into a resonance and viceversa, or
different resonances into each other. So the set of states
produced in the initial hard collision is furtherly  remixed by
FSI (this mechanism is similar to the  regeneration phenomenon
in the $K^o\overline K^o$ system). In the nuclear $N^*$
electroproduction, different Feynman graphs sum coherently as
shown in Fig. 1. In some kinematical regions the non-direct
contribution can dominate the process. For example, if we
measure $(e,e'N^*)$ at the quasi-elastic  peak of nucleon
production,  then the direct production of a 1535 MeV mass in
the $\gamma^* N$ collision is a very rare event, confined to the
tail at large $\vert \vec k \vert$ of  $\Phi(k)$. At the
quasi-elastic peak, a much stronger  possibility is given by a
two-step process where a nucleon is produced in the initial hard
scattering, and then converted by FSI into $N^*(1535)$.  At
other kinematics the easiest way to produce $N^*$(1535) can be
first to excite a higher  mass state (e.g. $S_{11}$(1650)) and
then to convert it into $N^*(1535)$ via FSI.

{}From now onwards the suffix $d$ represents the detected
``particle'' $X$ in $(e,e'X)$). The missing momentum  $\vec p_m
\equiv \vec p_d - \vec q$ is a parameter of $\Phi(k)$, and
consequently of $f(s)$: for $\vec k$ equal to $p_m$ one has  $s
= (q + k)^2 = M_d^2$.  Every value of $\vec k$ different from
$\vec p_m$ will lead  to some event that will not be {\sl
directly}\/ seen.  While $\vec p_m$ is measurable, $\vec k$ is
not measurable because it enters a loop (see Fig. 1).

If the detected particle is $N^*(1535)$, the curves in Fig.
2 represent the ($q,p_m$) pairs that lead to the largest direct
production of some states  with given mass. Parallel kinematics
($\vec p_m$ along $\vec q$) is  assumed in all figures. Positive
(negative) $p_m$ values correspond to $\vec p_m$ being parallel
(antiparallel) to $\vec q$. The $q$-axis is the maximum for
direct $N^*(1535)$ production. A DWBA calculation without final
remixing shows a significant $N^*$ production only in a narrow
band along the $q$-axis. But if we introduce the FSI remixing
with the nucleon channel, then we can have a two-step $N^*$
production in the region of negative $p_m$, around the maximum
for nucleon direct production. In the region of positive $p_m$
strong two-step contributions may come from the remixing with
the resonances of the third region. Unfortunately there are many
of them (see e.g. Ref. [2]) and their {\sl collective}\/
contribution is not easy to quantify if one does not know all
the amplitudes by which they enter the process.

For each intermediate channel $\vert i\rangle$ leading to a given
final state $\vert j\rangle$ we have to consider two amplitudes:
$B_i$ for  producing the intermediate state, and $f_{ij}$
associated with FSI  ($f_{ij}$ enter linearly into $V_{ij}$, see
Ref. [3]). The relative  phase and strength of the $i$-channel
contribution to  $\vert i\rangle$ production is given by $B_i
\cdot f_{ij}$. In the  regions where one channel (either direct
or not) dominates, the results  will not depend on the phase of
$B_i \cdot f_{ij}$.

At not too low energies the wave function of a state with mass
$M_i$ is approximately of the form  $D(\vec r) \cdot \exp(i \vec
p_i \cdot \vec r)$ (with $D(\vec r)$ slowly varying with $\vec
r$). For a given energy $E$, each $p_i$ is determined by  $p_i =
\sqrt{E^2 - M_i^2}$. So the coherence between the wave functions
in different channels is quickly lost if their masses are very
different. For this reason the mixing of $N^*(1535)$ with a
state, say, with M = 3 GeV is in practice much less important
than the mixing with a state at 1.7 GeV.

We work in a time-independent coupled-channel DWBA formalism,
with final state wave function formed by three coupled
components, labelled 1, 0 and  2, respectively: 1) direct
$N^*(1535)$, 0) nucleon, 2) a representative upper mass state
$N^*(1680)$. These are simultaneously produced with amplitudes
$B_i(Q^2,s)$ in the hard scattering at a given point $\vec r_o$.
A common phase $\exp(i\vec q \cdot \vec r_o)$ is given by the
virtual photon $\gamma^*$. Then we calculate a function of the
form $\exp(i\vec q \cdot \vec r_o) \cdot  \Sigma_i B_i
\Psi_i(\vec r,\vec r_o)$  ($\Psi_i(\vec r_o, \vec r_o) \equiv
1$), which is an eigenfunction  of the final state hamiltonian
$H_o + V$ (3x3 matrix). All the three channels describe
particles propagating in the same direction with the same
energy, and $p_i = \sqrt{E^2 - M_i^2}$. $V\equiv \{V_{ij},i,j=0,
1, 2\}$ is an optical potential describing FSI. Its nondiagonal
terms cause transitions between different channels. $V$ does not
conserve flux ($V^\dagger \neq V$), but each of its elements
conserves energy. The total wave function  $\Psi(\vec r)$ is  a
coherent sum of all the waves of the above kind emitted from any
nuclear point $\vec r_o$.

The details of the formalism applied to the coupled channel
problem can be found in  Ref. [3]. Some approximations have been
removed since then (Ref. [4]). Numerical calculations have been
carried out for $(e,e'N^*)$ on $^{27}Al$ with  shells up to
$l=2$ (harmonic oscillator wave functions).   $\vert B_i\vert$
is calculated as in Ref. [4]:

$$\vert B_i\vert^2 = \int d^3 k  \vert g_i(Q^2)\vert^2 R_i(s)
n(k), \eqno(1)$$
where $s \equiv (q+k)^\mu(q+k)_\mu$, $n(k)$ is a Woods-Saxon
distribution for the bound nucleon momentum $\vec k$,

$$R_i(s) = {{M_i\Gamma_i} \over {(s-M_i^2)+M_i^2\Gamma_i^2}}.
\eqno(2)$$
The function $g_i(Q^2)$ is the vertex form factor. Since only the
ratio $B_i/B_j$ is important here, all the $g(Q^2)$ are assumed
equal. This is correct in the $Q^2$ range where the quark
counting rule can be applied [5], and it is a good estimate at
$Q^2\sim$ some GeV$^2$.

The amplitude $B_i$ is a function of $\vec p_m \equiv \vec p_d -
\vec q$. Using four-momentum conservation in the hard vertex,
and energy conservation in the rescattering (which is correct in
the optical model), one can express $s$ (and so $\vert B_i
\vert$) as a function of $\vec q$ and $ \vec p_m$:

$$s - M_i^2 = 2\vec q \cdot (\vec p_m - \vec k) + p_m^2 -
k^2+M_d^2-M_i^2. \eqno(3)$$
The behaviour of the maxima of $\vert B_i \vert$ is shown in Fig.
3. Since the relevant phase here is that of the product $B_i
\cdot f_{ij}$, we absorb the phase of $B_i$ into $f_{ij}$.

All the nondiagonal FSI amplitudes $f_{ij}$ are chosen (in
modulus) as 1/2 the diagonal ones (all equal among them). This
can be estimated from the diffraction resonance production data
[6]. The phases are unknown to us. We have tested two opposite
possibilities:

\begin{itemize}
\item[$i$)] The relative phases of diagonal and nondiagonal
terms  are opposite (all equal inside each subgroup); this would
lead to the largest nuclear transparency at
$Q^2\rightarrow\infty$.

\item[$ii$)] All terms with the same phase (Fig. 4); this would
lead to the smallest nuclear transparency at
$Q^2\rightarrow\infty$.
\end{itemize}

The ratio $R$ between our results and PWBA is shown in Fig. 3
corresponding to the choice $i)$ for the phases of the
superposition amplitudes. The calculation for the choice $ii)$
gives the same results for large and negative $p_m$, thus
confirming that $N^*$ is mainly produced via nucleon convertion
in FSI. This contribution is locally  strong and independent of
the above phases. For  $p_m \gg 0$ we clearly  identify the
contribution by the upper resonance but, as noted above, we
cannot say much about the role played by many coherent
resonances. Anyway, unless many of them can sum up their
contributions, we espect the interference with the nucleon to
remain the most important. In parallel kinematics, this appears
at large  and {\sl negative}\/ $p_m$. Recalling eq. (3) we
notice that, at the leading order in $p_m/q$, $\vert
B_i\vert^2$  only depends on $\vec p_m$ via its component along
$\vec q$, and the sign is crucial. So this effect should be more
pronounced in the kinematics close to parallel.

There is a qualitative agreement between our results and the
observations in Ref. [1]: an enhancement of $N^*$ production
occurs at increasing absolute values of missing momentum. We
stress however that the experiment of Ref.  [1] did not
distinguish between positive and negative missing momenta.

In this sense we suggest a measure of the asymmetry

$$A \equiv
{{\sigma(\vec p_m\cdot\hat q > p_o) -
\sigma(\vec p_m\cdot\hat q < -p_o)} \over
{\sigma(\vec p_m\cdot\hat q > p_o) +
\sigma(\vec p_m\cdot\hat q < -p_o)}}, \eqno(4)$$
where a threshold $p_o$ is needed in order to exclude the central
region where the direct $N^*$ production is dominant. This
asymmetry is to large extent model independent and is equal to
zero in PWBA. It should be very sensitive to the suggested
interference effects.

\vskip 1.0cm

We are grateful to L.B.~Weinstein for useful discussions.

\clearpage

\vskip 0.5cm
\centerline{\Large{\bf Figure captions}}
\vskip 0.5cm

Fig. 1. The process of $N^*$ electroproduction off nucleus by
hard scattering.

Fig. 2. The maxima of the squared effective amplitudes $\vert
B_2\vert^2$ for the  $^{27}Al(e,e'p)$ reaction as a function of
four-momentum transfer $q$  (GeV) and missing momentum $p_m$ (in
units of Fermi momentum $p_F = 234$ MeV). The curves are labelled
according  to the mass of the intermediate baryonic resonance
mass.

Fig. 3. The transparency coefficient $R$ as a function of the
momentum transfer $q$ (in GeV) and the missing momentum $p_m$ (in
units of the Fermi momentum $p_f = 234$ MeV).

\end{document}